# Electron spin orientation dependence on momentum in pseudo gap phase of Bi2212

*Udai Prakash Tyagi and Partha Goswami(Retired)*

*Deshbandhu College, University of Delhi, Kalkaji, New Delhi-110019,India*

*E-mail of the first author: uptyagi@db.du.ac.in;*
*E-mail of the corresponding author: physicsgoswami@gmail.com;  pgoswami@db.du.ac.in.*

**ABSTRACT**

The aim of the paper is to study the electron spin direction dependence on momentum due to the presence of Rashba spin-orbit coupling in the pseudo-gap phase of Bi2212 bilayer. The non-trivial spin texture in **k**-space is found tuneable by electric field. The dependence is reported earlier by a group of workers in a spin- and angle-resolved photoemission spectroscopic measurement. The synthetic spin-orbit coupling, characterized by the broken time-reversal symmetry, is expected to be useful for the manipulation of the spin orientation.

## 1. Introduction

We investigate the band structure of $Bi_2Sr_2CaCu_2O_{8+\delta}$(Bi2212) bilayer system starting with a time reversal symmetry, and inversion symmetry broken Bloch Hamiltonian involving a term accounting for the effect of coupling between different $CuO_2$ planes. The coupling possesses a very different form in a single layer cuprate such as LSCO or NCCO than when a $CuO_2$ bilayer is present as in Bi2212. The Hamiltonian also involves Rashba spin-orbit coupling (RSOC). The origin of RSOC lies in the fact that whereas one Cu-O layer has Ca ions above and, Bi-O ions below, in the unit cell of the other layer this situation is reversed. This leads to a nonzero electric field within the unit cell. A recent spin- and angle-resolved photoemission spectroscopy measurement for $Bi_2Sr_2CaCu_2O_{8+\delta}$ had revealed **[1]** a non-trivial spin texture which corresponds to a well-defined direction for each electron real spin depending on its momentum. This spin-momentum locking (SML)suggests the presence of a strong spin-orbit coupling in Bi2212 like a topological insulator. The band structure analysis of this ambivalent system carried out in the paper, corresponding to the bonding and the anti-bonding cases, yields the following: The spin-momentum locking (SML) is evident, While the nodal region of the momentum space can give rise to a spin-down hole current, the anti-nodal region can give rise to spin-up electron current. We find that the common ground between two types of SMLs, corresponding to the bonding and the anti-bonding cases, is that the states of opposite spin are to be found in different parts of the Brillouin zone.

The conventional superconductors (SC) involve symmetric s-wave spin-singlet pairing of electrons by phonon-mediated attractive interaction, while the unconventional SCs require a long-range interaction [2] and have lower symmetry Cooper pairs. The Chiral SCs [3] are interesting instances of unconventional SCs where Cooper pairs with finite angular momentum circulate around a unique chiral axis. This leads to spontaneous breaking of time-reversal symmetry (TRS). The widely known member of this class is bulk $Sr_2RuO_4$, conjectured to be p-wave SC displaying spin- triplet pairing ; the chiral singlet SCs were believed to be very few and far between. Perhaps, the strongly correlated heavy fermion SC such as $URu_2Si_2$ [4] is a known example. The general belief regarding the featuring of spin-triplet p-wave pairing by the chiral SCs was actually based in the absence of a drop in NMR Knight shift [5], the broken TRS found in muon spin relaxation [6], and Kerr rotation experiments [7]. The NMR experiments [8-10] performed in the recent past, however, indicate utterly compelling irreconcilability with a number of odd-parity (pseudo-)spin triplet order parameters leading to the resurrection of the spin-singlet pairing scenario [11]. It must be mentioned that before these issues came to the fore, the nature of the pseudo-gap(PG) phase of the cuprate high temperature superconductors had already posed a number of unexplained puzzles. Below a characteristic temperature $T^*$ but higher than the SC transition temperature $T_c$, the excitation spectra showing a gap was first noticed by the relaxation rate of nuclear magnetic resonance [12] and then by many other transport and spectroscopic measurements [13]. But the most direct observation of this gap structure was shown by the ARPES [14]. The energy gap appears near the anti-nodal region, of the two-dimensional Brillouin zone (BZ) of the cuprate. The ARPES spectra at the anti-nodal region does not have the usual particle-hole symmetry associated with traditional superconductors. This asymmetric anti-nodal gap onsets at $T^*$ and it persists all the way to the SC phase [15,16]. There are four disconnected segments of Fermi surface near the nodal region. These segments called Fermi arcs, which are believed to be part of a small pocket [17, 18], have been reported to have their length not sensitive to temperature [19]. This presence of finite fraction of Fermi surface is consistent with the Knight shift measurement [20] showing a finite density of states (DOS) after the superconductivity is suppressed. Below $T_c$ the gap at anti-node merges with the SC gap. With the aim to explain the nature of PG, many reports have appeared in not so distant past, where charge density waves (CDW) or spin density waves (SDW) were related to SC and PG phases [21-28]. On another level, the pseudo-gap is a distinct phase akin to an unconventional metal or, a symmetry preserved/broken state [29-41]. The onset of the pseudo-gap is defined by the opening of an anti-nodal gap and the reduction of the large Fermi surface to Fermi arc(area in momentum space that remains un-gapped) [17-19,29-41] as in Wyle/Dirac semi-metal. The exact nature of the pseudo-gap still remains an open issue.

In the past, Nayak et al.**[ 33-35 ]** have put forward *d* density wave (DDW) order as the one for the pseudo gap (PG)phase of the hole-doped cuprates. This order [33-35] corresponds to spontaneous currents along the bonds of a square lattice for the ordering wave vector $\mathbf{Q}=(\pm\pi,\pm\pi)$. The DDW state preserves the combined effect of the microscopic time reversal symmetry violation and the translation by a lattice spacing. The net result is that the staggered magnetic flux produced by these currents is zero on the macroscopic scale. The formulation of Nayak et al.**[33-35]** involved d-wave superconductivity(DSC). The reason for the choice is that the d-wave superconducting symmetry and strong Coulomb repulsion are compatible :For strong on-site Coulomb repulsion the superconducting state needs to avoid same-site pairing which corresponds to isotropic k-space pairing. At the same time, the Coulomb repulsion generally favours anti-ferromagnetic tendencies and thus a spin-singlet superconducting state. The spin singlet state with the lowest number of nodes but still avoiding same-site pairing is exactly the d-wave state. The non-zero DSC gap $\Delta_k^{(sc)}$ requires an appropriate attractive interaction. The gap is, formally similar to the weak- coupling BCS gap equation, given by $\Delta_k^{(sc)}=\Delta_0(T)(\cos k_x a - \cos k_y a)$. The Chiral DDW (CDDW) order, which is a complex variant of the DDW order, was subsequently advocated for by S. Das Sarma et al. **[36-39]** as the one suitable for the PG phases ordering. The reason being it offers a theoretical explanation [36-39] of the non-zero polar Kerr effect observed in YBCO by Kapitulnik et al.**[39]**. A very pertinent theoretical reason for the identification of the PG state with the CDDW state rather than with the DDW state is that the imaginary part of the d+id-wave order parameter $D_k=(-\chi_k+i\Delta_k)$ breaks the parity and the time reversal symmetry (TRS) of the normal state. The quantities $(\chi_k, \Delta_k)$ are given by **[36-39]** $\chi_k= -(\chi_0/2) \sin(k_x a) \sin(k_y a), \Delta_k= (\Delta_0^{(PG)}(T)/2)(\cos k_x a - \cos k_y a)$, and $\mathbf{k} = (k_x, k_y)$. The normal state dispersion is imperfectly nested here. The corresponding evidence is that the onset of CDDW ordering leads to a peak in the anomalous Nernst signal (ANS)[36,37]. The main contribution to this chirality induced ANS comes from the points $(\pm\pi(1-\varphi), \pm\pi\varphi),(\pm\pi\varphi, \pm\pi(1-\varphi))$ with $\varphi \sim 0.2258$ located roughly on the boundary of the Fermi pockets in the momentum space( cf. ref. **[36,37]**). Thus, in order to present a suitable description of cuprates in the framework of CDDW ordering, we have to turn our attention to this imperfectly nested dispersion. All energies are expressed in units of the first neighbour hopping. The second-neighbour hopping in the dispersion, which is known to be important for cuprates**[33-44]**, frustrates the kinetic energy of electrons.

The paper is organized as follows: In section 2 we derive an expression for the single-particle excitation spectrum in CDDW state. We show the spin-momentum locking (SML) in section 3. We also investigate the spin texture here. We discuss about Dzyaloshinskii–Moriya interaction (DMI) which will be useful in a sequel to this work. The paper ends with brief discussion ( mainly on the possible deployment of synthetic spin-orbit coupling for the manipulation of the spin orientation) and conclusion in section 4.

## 2. Excitation spectrum with interlayer tunnelling and Rashba coupling

Motivated by the findings of S. Das Sarma et al.**[36-39]**, we assume that the CDDW order represents the pseudo-gap(PG) phase. We assume the momentum dependence of the pairing interactions required for this kind of ordering is given by the functions of the form $U_{x2-y2}(k,k') = U_1 (\cos k_x a - \cos k_y a)(\cos k'_x a - \cos k'_y a)$, $U_{xy}(k,k')= U_2 \sin(k_x a) \sin(k_y a) \sin(k'_x a) \sin(k'_y a)$, where $U_1$ and $U_2$ ($U_1 > U_2$) are the coupling strengths, and $(k_x,k_y)$ belong to the first Brillouin zone (BZ). Furthermore, the cuprates (the special class of high-temperature superconductors (HTSCs)) consist of superconducting $CuO_2$ layers separated by spacer layers as in Bi2212. The simplest vertical hop is straight up via $t_b$. The bilayer split bands in Bi2212 display considerable dispersion with k: It corresponds to $[-2t_b(c_x(a) - c_y(a))^2]$ where $t_b$ is an effective parameter for hopping within a single bilayer, i.e. it controls the intracell bilayer splitting and $c_j(a) \equiv \cos(k_j a)$ **[45]**. The compound Bi2212, however, involves intercell hopping ($t_z$), too. The total dispersion can now be written as $\epsilon(\mathbf{k},k_z) = \varepsilon_\mathbf{k} - \mu + \varepsilon_{k_z}(\mathbf{k})$, where $\mathbf{k}$ and $k_z$ respectively denote the in-plane and out-of-plane components of $\mathbf{K} = (\mathbf{k},k_z)$. The dispersion $\varepsilon_\mathbf{k}$ has the usual form

$$\varepsilon_k = -2t\left(c_x(a) + c_y(a)\right) + 4t'c_x(a)c_y(a) - 2t''\left(c_x(2a) + c_y(2a)\right) - 4t'''(c_x(a) c_y(2a)$$

$$+c_y(a) c_x(2a)) \quad (1)$$

including the first, the second and the third neighbour hops due to the non-zero neighbour hopping and μ is the chemical potential of the fermion number. For simplicity, we have set the lattice constant equal to one. The term $\varepsilon_k$ is the model dispersion associated with a single CuO$_2$ plane if the effects of $k_z$-dispersion are totally neglected. Early theories only took t into account, but the consistent results of local-density approximation, band-structure calculations **[17-19]** and angle-resolved photoemission spectroscopy (for over-doped, stripe-free materials) **[17-19]**, have led to the usage of including also t', with t'/t = 0.1 for La$_2$CuO$_4$ and t'/t = 0.3 for YBa$_2$Cu$_3$O$_7$ and Bi$_2$Sr$_2$Ca Cu$_2$O$_8$, whereby the constant-energy contours of the expression for $\varepsilon_k$ become rounded squares oriented in the [11]- and [10]-directions respectively. For the hole-doped materials, t′ > 0 (for the electron-doped materials t′ < 0), and, in all cases, t′ < (t/2). The $\varepsilon_{k_z}(k)$ term accounts for the effect of coupling between different CuO$_2$ planes, and possesses a very different form in a single layer cuprate such as LSCO or NCCO than when a CuO$_2$ bilayer is present as in Bi2212. In the case of Bi2212**[46]**,

$$\varepsilon_{k_z}(k) = -\Gamma_z(k, c_z(c/2))[(c_x(a) - c_y(a))^2 /4 + a_0] \quad (2)$$

where c denotes the lattice constant along the z-axis, and

$$\Gamma_z(k, c_z(c/2)) = \pm (t_b^2 + A'^2 + 2t_b A' c_z(c/2))^{1/2}, A' = 4 t_z c_x(a/2) c_y(a/2). \quad (3)$$

The plus (minus) sign refers to the bonding (anti-bonding) solution and the term $t_z$ corresponds to the intercell hopping. The term $c_z$ arises because supposedly we have an infinite number of stacked layers. The term $A' = 4 t_z c_x(a/2) c_y(a/2)$ is zero along the high symmetry line X(π,0) —M(π, π). This leads to a lack of $k_z$-dispersion along this high-symmetry line. The additional hopping $a_0$ allows for the presence of a splitting at Γ(0,0). It is reported **[47]** that adequate control of the interlayer spacing albeit the interlayer hopping in Bi-based superconductors is possible through the intercalation of guest molecules between the layers. This could be a way to tune the hopping parameter. The coupling $t_b$ within a bilayer and the intercell coupling $t_z$ are both quite substantial as are the in-plane hopping terms beyond the NN term. It may be mentioned that Vishwanath et al.**[1]** had found that a spin-momentum locking that allows states of opposite spin to be localized in different parts of the unit cell. The spin-momentum locking, to be obtained in this section, which is similar to their observation in spirit save for the fact that the Wigner-Seitz cell is replaced by the Brillouin zone (BZ). We shall see that when the Rashba-coupling is introduced together with the full form of the tunnelling matrix given by Eq.(2) and (3), while the nodal region of the momentum space will give rise to a spin-down hole current, the anti-nodal region will give rise to spin-up electron current in the bonding case. In the anti-bonding case, the result will be similar.

Suppose now $d^{(m)}_{k,\sigma}$ ( σ = ±1 for the real spin ) corresponds to the fermion annihilation operator for the single-particle state (**k**,σ) in the layer m (m=1,2) of the system. As regards the interactions, $U \sum_i d^\dagger_{i\uparrow} d_{i\uparrow} d^\dagger_{i\downarrow} d_{i\downarrow}$ is the onsite repulsion of *d* electrons, where the intra-layer *d* electrons are locally interacting via a Hubbard-$U$ repulsion. We have not considered this term assuming the correlation effect are marginally relevant. In the basis $(d^{\dagger(1)}_{k,,\sigma}, d^{\dagger(1)}_{k,+Q,\sigma}, d^{\dagger(2)}_{k,-\sigma}, d^{\dagger(2)}_{k+Q,-\sigma},)^T$, we consider the following bilayer Hamiltonian in momentum space :

$$H(k_x, k_y, k_z) = \begin{pmatrix} \varepsilon_k & D^{\dagger}_k & \tau(\mathbf{k}, k_z) & 0 \\ D_k & \varepsilon_{k+Q} & 0 & 0 \\ \tau^*(\mathbf{k}, k_z) & 0 & \varepsilon_k & D^{\dagger}_k \\ 0 & 0 & D_k & \varepsilon_{k+Q} \end{pmatrix} \quad (4)$$

where $\tau(\mathbf{k}, k_z) = \varepsilon_{k_z}(\mathbf{k}) + \alpha_k$, $\alpha_k = \alpha_0(-i\sin(k_x a) - \sin(k_y a))$ is the polarizing field(assumed to be in the z direction) led(Rashba) spin-orbit coupling. The quantity $\alpha_0$ is the coupling strength which is proportional to the strength of the field. Therefore, it is tunable. Here, as before, we assume the ordering wave vector $\mathbf{Q} = (\pm\pi, \pm\pi)$. We shall also consider the case $\mathbf{Q} = (\pm Q_1, \pm Q_2)$ where ($Q_1 = 0.7742\pi$, $Q_2 = 0.2258\pi$). The occurrence of spin-flip is due to Rashba coupling $\alpha_k$. Together with the lifting of the spin-degeneracy, indication of spin-momentum locking is then expected.

We find from above that $H(k_x, k_y, k_z)$ is inversion asymmetric. The reason for the broken mirror reflection symmetry, apart from the presence of the Rashba coupling, is $D_k(k_x \to -k_x, k_y \to k_y) \neq D_k(k_x, k_y)$ due to the presence of chirality(d+id). For the non-chiral system(id), however, $D_k(k_x \to -k_x, k_y \to k_y) = D_k(k_x, k_y)$, i.e. the inversion symmetry is preserved. Note that the Hamiltonian in (4) violates time reversal symmetry(TRS) also. Never-the-less, as we see below, we get real eigenvalues in BZ. Upon ignoring $\tau(\mathbf{k}, k_z)$ altogether, the energy eigenvalues in the CDDW case are given by multiple roots. These are $E^{(\alpha)}(\mathbf{k}) = \varepsilon_k^U + \alpha\Delta_k$ where $\alpha = \pm 1$, $\varepsilon_k^U = (\varepsilon_k + \varepsilon_{k+Q})/2$, $\varepsilon_k^L = (\varepsilon_k - \varepsilon_{k+Q})/2$, and $\Delta_k = \left(\varepsilon_k^{L^2} + |D_k|^2\right)^{\frac{1}{2}}$. The plots of free electron dispersion and, once the CDDW order sets in. are shown in Figure1. We find that the valence and the conduction bands corresponding to free electrons are partially full(see 1(a) and 1(b)). Therefore, the system conducts. Once the CDDW order sets in, the pseudo-gap phase displays a nodal-anti-nodal dichotomous feature, i.e. excitations with infinite lifetimes(say, in a Hartee-Fock treatment of the e-e repulsion), have un-gapped nodal points and maximally gapped anti-nodal points (see 1(c) and (d)). The reason basically lies in the particle-hole asymmetry of the excitation spectrum $\varepsilon_1, \varepsilon_2 = \varepsilon_k^U \pm \sqrt{\varepsilon_k^{L^2} + D_k^{\dagger}D_k}$, $\varepsilon_k^U = \frac{\varepsilon_k + \varepsilon_{k+Q}}{2}$, and $\varepsilon_k^L = \frac{\varepsilon_k - \varepsilon_{k+Q}}{2}$. With $Q_1 = 0.7742$pi, $Q_2 = 0.2258$pi, $\varepsilon_k^U \sim 0$, and $\varepsilon_k^L$ is minimum (maximum) at the nodal(antinodal) point.

In the case of the perfectly nested dispersion ($\varepsilon_k = -\varepsilon_{k+Q}$), it is easy to see that the eigenvalue equation which is a quartic may be written as $\varepsilon^4 - 2\varepsilon^2 b - 4\varepsilon c + d = 0$ or,

$$\varepsilon^4 - 2\varepsilon^2 b + b^2 = 4\varepsilon c + b^2 - d, \quad (5)$$

where $b = \frac{1}{2}(t_k^2 + 2(\varepsilon_k^2 + |D_k|^2))$, $c = \frac{1}{2}\varepsilon_k t_k^2$, and $d = (\varepsilon_k^2 + |D_k|^2)^2 - t_k^2\varepsilon_k^2$. For simplicity, we have replaced here $\tau(\mathbf{k}, k_z)$ by $t_k$. We now add and subtract an as yet unknown variable z within the squared term $(\varepsilon^4 - 2\varepsilon^2 b + b^2)$: $(\varepsilon^2 - b + z - z)^2 = 4\varepsilon c + b^2 - d$, or $(\varepsilon^2 - b + z)^2 = 2z\varepsilon^2 + 4\varepsilon c + (z^2 - 2bz + b^2 - d)$. Upon ignoring the term $(4\varepsilon c)$ we shall get a bi-quadratic with values of '$\varepsilon$' given by $\varepsilon^2 \approx b \pm \sqrt{(b^2-d)}$. We shall see below that without the term $(4\varepsilon c)$ a discussion of the spin-texture, etc., which is one of our tasks here, does not seem to be possible. The left-hand side of Eq.(5) is a perfect square in $\varepsilon$. Therefore, we need to rewrite the right hand side in that form as well. For this we require that the discriminant of the quadratic in the variable $\varepsilon$ to be zero. This yields $16c^2 - 8z(z^2 - 2bz + b^2 - d) = 0$ or, $z^3 - 2bz^2 + (b^2 - d)z - 2c^2 = 0$. The corresponding depressed cubic equation $s^3 + ps + q = 0$ has the discriminant function $D = -4p^3 - 27q^2$, where $s = z - (2b/3)$, $p = -(b^2/3 + d)$, and $q = 2(b^3/27 - bd/3 - c^2)$. Since we find $D$ negative in the entire Brillouin zone, the equation $s^3 + ps + q = 0$ has one real root and two complex conjugate roots. Suppose we denote this root by $s_0(a\mathbf{k})$, then the corresponding '$z$' will be denoted by $z_0(a\mathbf{k}) = s_0(a\mathbf{k}) + 2b/3$. After lengthy algebra we obtain real $z_0 = 2b/3 + (-q/2 + 0.5\psi^{1/2})^{1/3} + (-q/2 - 0.5\psi^{1/2})^{1/3}$, where $\psi = q^2 + 4(b^3/9 + d/3)^3$. Using (5) one may then write $\varepsilon^2 = b - z_0 \pm \{\sqrt{(2z_0)}\,\varepsilon + \sqrt{(2/z_0)}c\}$ or, $\varepsilon^2 \mp \sqrt{(2z_0)}\,\varepsilon + (-b + z_0 \mp c\sqrt{(2/z_0)}) = 0$. These equations are expected to yield the band structure. We, however, also wish to investigate the consequence of the imperfectly nested dispersion. In what follows we, therefore, proceed in a straightforward manner without any assumption.

The eigenvalues of the matrix in (4) are, once again, given by the quartic $A\varepsilon^4 + B(k)\varepsilon^3 + C(k)\varepsilon^2 + D(k)\varepsilon + E = 0$, where

$$A = 1, \quad B(k) = -4\varepsilon_k^U, \quad C(k) = 4\varepsilon_k^{U^2} + 2\gamma_1^2 - \gamma_2^2 - \gamma_3^2,$$

$$D(k) = -4\varepsilon_k^U \gamma_1^2 + 2\varepsilon_{k+Q}\gamma_2^2 + 2\varepsilon_k\gamma_3^2, \quad E(k) = \gamma_1^4 - \gamma_2^2 \varepsilon_{k+Q}^2 - \gamma_3^2 \varepsilon_k^2 + \gamma_0^4,$$

$$\varepsilon_k^U = \frac{\varepsilon_k + \varepsilon_{k+Q}}{2}, \quad \varepsilon_k^L = \frac{\varepsilon_k - \varepsilon_{k+Q}}{2}, \quad \gamma_1^2 = (\varepsilon_k \varepsilon_{k+Q} - |D_k|^2),$$

$$\gamma_2^2 = \{(\varepsilon_{k_z}(\mathbf{k}) - \alpha_0 \sin(k_y a))^2 + \alpha_0^2 \sin(k_x a)\},$$

$$\gamma_3^2 = \{(\varepsilon_{k_z}(\mathbf{k+Q}) - \alpha_0 \sin(k_y a + Q_2))^2 + \alpha_0^2 \sin(k_x a + Q_1)\},$$

$$\gamma_0^4 = -2|D_k|^2(\varepsilon_{k_z}(\mathbf{k}) - \alpha_0 \sin(k_y a))(\varepsilon_{k_z}(\mathbf{k+Q}) - \alpha_0 \sin(k_y a + Q_2))$$

$$-2|D_k|^2 \alpha_0^2 \sin(k_x a + Q_1) \times \sin(k_x a) + \gamma_2^2 \gamma_3^2. \quad (7)$$

In view of the Ferrari's solution of a quartic equation, we find the roots as

$$\epsilon_j(s, \sigma, k) = \sigma \sqrt{\frac{z_0(k)}{2}} + \varepsilon_k^U + s\left(b_0(k) - \left(\frac{z_0(k)}{2}\right) + \sigma c_0(k)\sqrt{\frac{2}{z_0(k)}}\right)^{\frac{1}{2}}, \quad (8)$$

where $j = 1, 2, 3, 4$, $\sigma = \pm 1$ is the spin index and $s = \pm 1$ is the band-index. The other functions appearing in (3) are defined below:

$$z_0(k) = \frac{2b_0(k)}{3} + \left(\frac{1}{2}\Delta^{\frac{1}{2}}(k) - A_0(k)\right)^{\frac{1}{3}} - \left(\frac{1}{2}\Delta^{\frac{1}{2}}(k) + A_0(k)\right)^{\frac{1}{3}}, (9)$$

$$A_0(k) = \left(\frac{b_0^3(k)}{27} - \frac{b_0(k)d_0(k)}{3} - c_0^2(k)\right), b_0(k) = \frac{3B^2(k) - 8C(k)}{16}, c_0(k) = \frac{-B^3(k) + 4B(k)C(k) - 8D(k)}{32}, \quad (10)$$

$$d_0(k) = \frac{-3B^4(k) + 256E(k) - 64B(k)D(k) + 16B^2(k)C(k)}{256}, \quad (11)$$

$$\Delta(k) = \left(\frac{8}{729}b_0^6 + \frac{16d_0^2 b_0^2}{27} + 4c_0^4 - \frac{4d_0 b_0^4}{81} - \frac{8c_0^2 b_0^3}{27} + \frac{8c_0^2 b_0 d_0}{3} + \frac{4}{27}d_0^3\right), \quad (12)$$

One can gap out the helical edge states by introducing a Zeeman term that explicitly breaks the protecting time-reversal symmetry. As we have seen above in Eq. (8), we obtain a term $\sqrt{\frac{z_0(k)}{2}}$ which has different sign for opposite spins and connection with momentum (see Figure-3).Usually, the effects of an external magnetic field, B, perpendicular to the CuO$_2$plane may be captured by two additional terms in the planar Hamiltonian. The first term describes the orbital coupling to magnetic field through the minimal coupling k$_x$ → k$_x$ + (e/c)A$_x$, where one may choose **A** = [−By, 0] is the vector potential in the Landau gauge. The second term describes the coupling of the spins to the magnetic field and is given by the Zeeman contribution. Since, the spin index $\sigma$ occurs twice in Eq. (8)(See $\epsilon_j(s,\sigma,k) = \sigma\sqrt{\frac{z_0(k)}{2}} + \varepsilon_k^U + s\left(b_0(k) - \left(\frac{z_0(k)}{2}\right) + \sigma c_0(k)\sqrt{\frac{2}{z_0(k)}}\right)^{\frac{1}{2}}$, where the index is highlighted by red ink.), the term $\sqrt{\frac{z_0(k)}{2}}$in question does not act like magnetic energy.

## 3. Spin-momentum locking and spin texture

The plots of bands in (8) in the bonding/ anti-bonding cases belonging to the nodal and anti-nodal regions are shown in Figures 2. The numerical values of the parameters used in the plots in Figure 2 are t =1 , t′/t = −0.28(hole-doping), t ″/t = 0.1, t ‴/t = 0.06, t$_b$/t = 0.3, t$_z$/t = 0.1, $\frac{\alpha_0}{t} = 0.53$, and a$_0$ = 0.4. Throughout the whole paper, we choose ***t*** to be the unit of energy. In Figure 2(a) we have a plot of quasi-particle excitation (QP) spectrum given by Eq.(8) in the bonding case as function of dimensionless momentum k$_x$a for k$_y$a = π/2 and k$_z$a = π. Since spin-down valence and conduction bands are partially empty , the spin-down QP conduction is possible. A plot of QPsin the bonding case, as a function of dimensionless momentum k$_x$a for k$_y$a = π and k$_z$a = π is given in Figure (b).  The plot displays band crossing and huge spectral gap at the high symmetry point R(π,π,π).A plot of quasi-particle excitation spectrum in the anti-bonding case as function of k$_x$a for k$_y$a = 0 and k$_z$a = 0 is shown in Figure (c).  Since the spin-up conduction band is partially empty, the spin-up electron conduction is possible. A band pair of opposite spins, where one of the partners is a partially empty band and the other is full band, is almost coinciding in energy with zero density of states(DOS) at their meeting point in momentum space. In Figure 2(d) we have a plot of QP excitation spectrum in the anti-bonding case as function of dimensionless momentum k$_x$a for k$_y$a = π/2 and k$_z$a = 0. Once again, since the spin-down valence band is partially empty ,

the spin-down hole conduction is possible. Thus, the spin-momentum locking (SML) is evident, While the nodal region of the momentum space can give rise to a spin-down hole current, the anti-nodal region can give rise to spin-up electron current. We find that the common ground between two types of SMLs, corresponding to the bonding and the anti-bonding cases, is that the states of opposite spin are to be found in different parts of the Brillouin zone. On the experimental front, recently Vishwanath et al.[1] discovered that Bi2212, has a nontrivial spin texture with the spin-momentum locking. They used spin- and angle-resolved photoemission spectroscopic technique to unravel this fact. We shall discuss briefly below this feature.

The spin texture of the surface states in topological insulator(TI) forms due to SOC. For the Bi2212 system also it is expected that Rashba SOC will induce this texture. Therefore, we now focus on this aspect. The spin texture $s(n,k)$ is defined as the expectation value of a vector operator $S_j =$ $I_{2\times 2} \otimes \sigma_j$ where $\sigma_j$ are Pauli matrices on a two dimensional $k$-grid and $\otimes$ stands for the tensor product. At $k$ for the state n (or nth band) it is defined as an expectation value $s_z(n,k)$ = $\langle S_z \rangle^{(n)} = \langle n|S_z|n \rangle$. Obviously enough, to calculate this we need eigenvectors of the Hamiltonian matrix (4) for an eigenvalue $E_n$. This complete set is given by

$$|\Psi_n\rangle = \varsigma_n^{-1}(\mathbf{k}, k_z) \begin{pmatrix} 1 \\ \frac{D_k}{E_n - \varepsilon_{k+Q}} \\ \frac{\tau^*(\mathbf{k},k_z)(E_n - \varepsilon_{k+Q})}{[(E_n - \varepsilon_{k+Q})(E_n - \varepsilon_k) - |D_k|^2]} \\ \frac{\tau^*(\mathbf{k},k_z) D_k}{[(E_n - \varepsilon_{k+Q})(E_n - \varepsilon_k) - |D_k|^2]} \end{pmatrix}, \quad n = 1,2,3,4, \qquad (13)$$

$$\varsigma_n(\mathbf{k}, k_z) = \sqrt{\left(\frac{(|D_k|^2)}{(E_n - \varepsilon_{k+Q})^2} + \frac{(|\tau(\mathbf{k},k_z)|^2)(E_n - \varepsilon_{k+Q})^2}{(E_n - \varepsilon_1)^2 (E_n - \varepsilon_2)^2} + \frac{(|\tau(\mathbf{k},k_z)|^2)(|D_k|^2)}{(E_n - \varepsilon_1)^2 (E_n - \varepsilon_2)^2}\right)} \qquad (14)$$

$$\tau(\mathbf{k}, k_z) = \varepsilon_{k_z}(\mathbf{k}) + \alpha_k, \alpha_k = \alpha_0(i\sin(k_x a) - \sin(k_y a)) \qquad (15)$$

where the full expression for $\varepsilon_{k_z}(\mathbf{k})$ could be found in Eqs.(2) and (3). Upon using (13)-(15), the spin textures are obtained in a straightforward manner. The contour plots are shown in Figure 4 for various values of $a_0$ and $\alpha_0$ with $ak_z = \pi$. The texture is found over a large range of parameters in this material with Rashba SOC and broken surface inversion starting with a critical value of $\alpha_0$. We have found this critical value to be 0,35 for $a_0 = 0.40$. For $\alpha_0 <$ 0.35, the expectation value turns out to be complex. Most importantly, as can be seen from Figure 4 that, the texture or the expectation value spreads over the larger region of the Brillouin zone as $\alpha_0$ (and $a_0$) increases. The texture is tunable by electric field(and by intercalation[47]). Upon using (13)-(15) we also obtain the values of ($\langle n|S_x|n \rangle$ - $\langle n|S_y|n \rangle$)spin texture in the bonding case in the $ak_x$−$ak_y$ plane for $ak_z = \pi$(see Figure 5) with the polarising field $\alpha_0$ in the z direction. The plots are for different values of $a_0$ and $\alpha_0$.

The Dzyaloshinskii–Moriya interaction (DMI) energy **[48-51]** is considered as one of the most important energies for specific chiral textures such as magnetic skyrmions. These are particle-like solitonic, complicated spin textures of topological origin existing in the real space. The interaction was introduced first by Stevens**[49]** where it was derived as a consequence of the inclusion of spin-orbit coupling in the Heisenberg model. In order to explain the weak ferromagnetic(FM) moments in largely anti-FM α-$Fe_2O_3$ crystals, the DMI was later introduced by Dzyaloshinskii**[50]**. The inclusion of spin-orbit coupling in a consideration of the super-exchange mechanism to provide the first microscopic derivation was reported by Moriya**[51]** later. He introduced a symmetric tensor of second order in the spin-orbit coupling and the Moriya vector **$D_{ij}$** which is proportional to the first power of the spin-orbit coupling and an antisymmetric vector. The DMI **[48]** is considered as one of the most important energies for specific chiral textures such as magnetic skyrmions. The keys of generating DMI are the absence of structural inversion symmetry and energy corresponding to spin–orbit coupling. An investigation of the Moriya vector **$D_{ij}$** in similar lines as that of Moriya**[51]** starting with the Bloch Hamiltonian for Bi2212 bilayer system will be carried out in a separate communication.

## 4. Discussion and Conclusion

The Hamiltonian in ref. [36,37] is inversion symmetry protected (no Rashba spin-orbit coupling(RSOC))as long as the ordering wave vector **Q**=($\pm\pi,\pm\pi$). The authors in this reference have claimed the possibility of quantum anomalous Hall (QAH)effect. The reason being a system which is inversion symmetry protected but time reversal symmetry (TRS) broken will have non-zero Berry curvature (BC).This is very much required to obtain anomalous Hall conductivity ($\sigma_{xy}$) and to show that $\sigma_{xy}$ is quantized in the case of an insulator. In this backdrop, an important question then arises here : "what is the necessity of breaking inversion symmetry (IS)"?The nontrivial spin texture could also be explained without it [52].In their seminal work **[52]**, Zhang et al. have demonstrated that the lack of the local inversion symmetry at atomic sites leads to hidden spin polarization completely determined by the site-dependent orbital angular momentum even in centrosymmetric crystals. This was an outcome of the first-principles calculations by the authors. Usually, the reason behind this is the orbital magnetization being more important than the spin magnetization, i.e. the spin–orbit coupling (SOC) is weak, such as a transition metal dichalcogenide $MoS_2$. Additionally, Baidya et al. **[53]** had also explored the important role of orbital polarization in QAH phases. A comparison with DFT calculation led these authors to the conclusion that effect of such terms are smaller. For Bi2212 system, however, we find that the spin texture demonstrates a momentum-selective spin polarization in a significant portion of the Brillouin zone, and the texture disappears when RSOC tends towards zero. Therefore, for our system, to have access to nontrivial spin texture broken inversion symmetry is needed.

Efforts have been made in the past for efficient spin generation and/or detection by controlling the flow of charge or spin currents **[54,55]** in other systems where the spin orientation is forced to align perpendicularly to the electron momentum. Such generation and detection process constitute important building blocks for future topological electronics and spintronics. To this end, a full set of spin controls based on SML spin manipulation is needed. Since the spin orientation is regulated by the electron momentum direction, the manipulation of the spin orientation, say, under the drift and diffusion makes the task difficult. In fact, this is a serious impediment to the potential scope of SML to provide flexibility in the design of spintronic devices.

The synthetic spin-orbit coupling (s-SOC)[56-59] arising due to Zeeman Hamiltonian involving the position-dependent magnetic field ( $B_0$ ) may prove to be useful for the spin manipulation. It must be noted that while s-SOC breaks time reversal symmetry (TRS), the intrinsic spin-orbit coupling (i-SOC) comprising of the Rashba and Dresselhaus SOCs (where the former is due to structural inversion asymmetry and the latter is due to bulk inversion asymmetry) are time-reversal symmetric. As the first step to show theoretically the possibility of the manipulation of the spin orientation in Bi2212 bilayer system by s-SOC, we need to obtain spin Hamiltonian, from the system Hamiltonian, by performing an Schrieffer–Wolff (SW) transformation. Next, we need to consider an oscillating electric field $E(t)$, applied, say, along the x-axis for the spin resonance purpose. Driven by this electric field, the electron spin experiences an effective oscillating magnetic field via the s-SOC. The Hamiltonian of the system is then expected to be comprising of transverse and longitudinal magnetic field components. In contrast to the i-SOC-mediated spin resonance **[58],** due to the broken TRS in the case of s-SOC, the transverse effective field is expected not to depend on the magnitude $B_0$ **[57].** If this happens we have hit the bull's eye. The reason being a weak $B_0$ will be sufficient in order to achieve the objective without sacrificing the speed of the electric-dipole spin resonance (EDSR). The charge-noise-induced spin dephasing mediated by s-SOC could also be supressed by a suitable choice of the magnetic field direction. The details will be presented in a sequel to this work.

In conclusion, the presence of Rashba spin–orbit coupling (possibly s-SOC as well) plays a key role on various aspects of spin transport in various systems. It can be detected in a material via magnetic field-induced quantum oscillations, EDSR, and weak antilocalization. In the context of the present problem, we have witnessed the role it has played. To have access to nontrivial spin texture non-zero RSOC(broken inversion symmetry) is needed. The potentially important role of hidden orbital polarizations[52,53] needs to be investigated deeply . The complete investigation of the role of s-SOC in the context of the spin orientation manipulation is expected to unravel more of it. There are other problems, such as the Zitterbewegung effect, quantum anomalous and magnetoelectric effects, and Floquet physics, which need our attention.

## Figures and Figure Captions

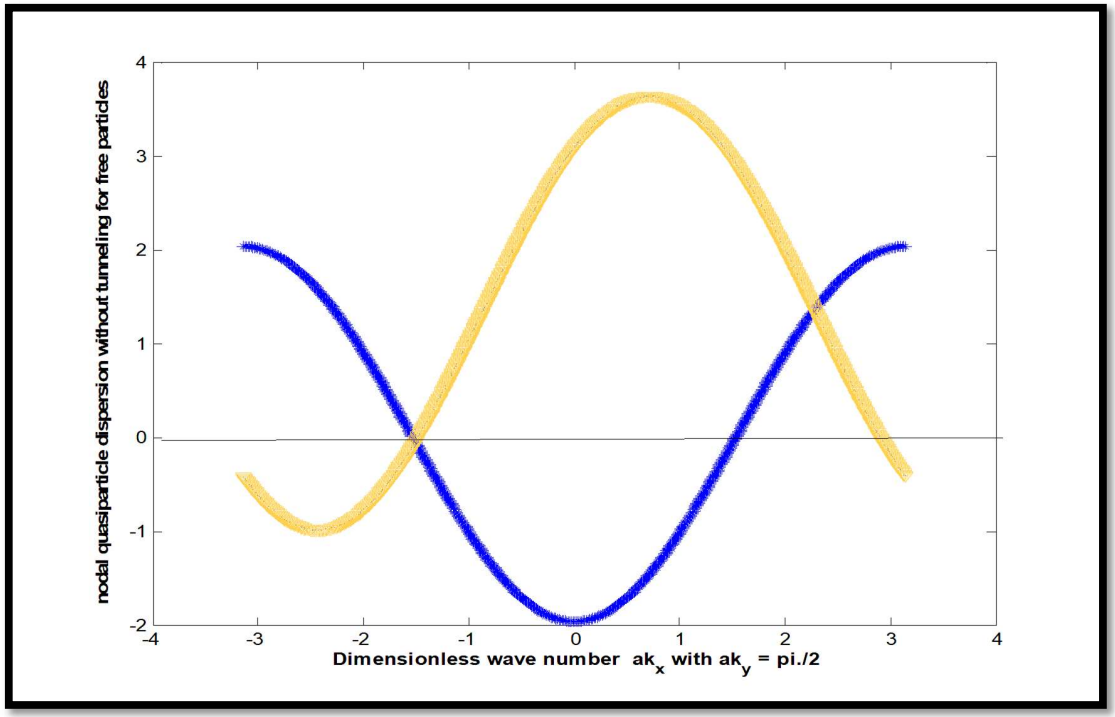

(a)

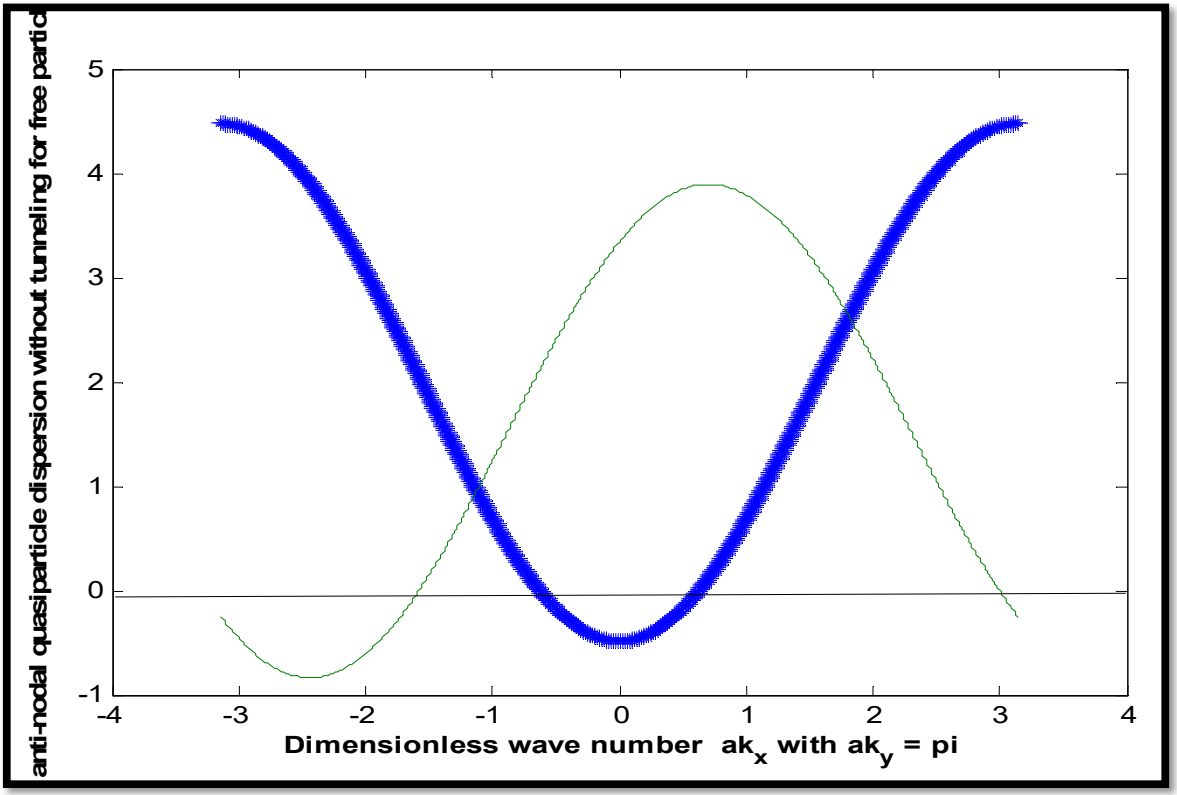

(b)

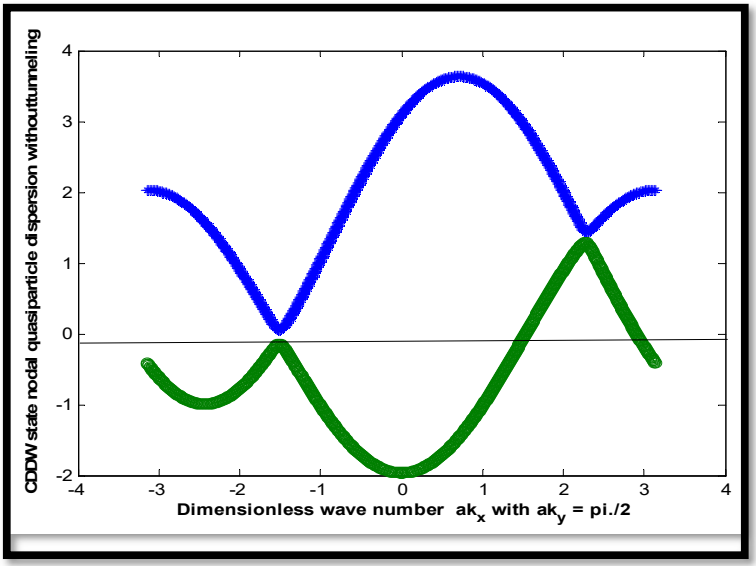

(c)

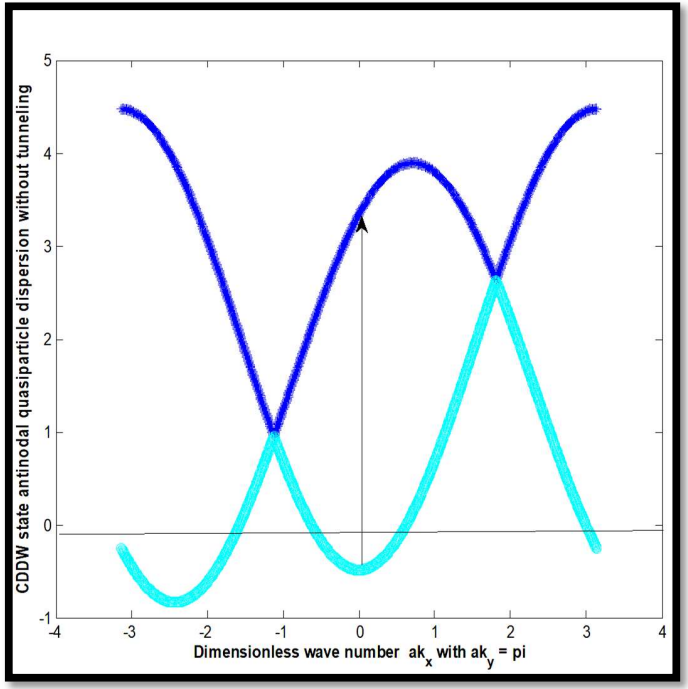

(d)

**Figure 1.** The two-band energy spectra of free electron dispersion and CDDW state ($\varepsilon_1, \varepsilon_2 = \varepsilon_k^U \pm \Delta_k$) without and with the interlayer tunneling at the nodal($ak_x \sim \frac{\pi}{2}, ak_y = \frac{\pi}{2}$) and the anti-nodal($ak_x \sim 0, ak_y = \pi$) regions. The chemical potential µ represented by solid, horizontal line µ ~0 is located as shown. In figures (a) and (b), the free particle bands are partially empty and hence conduction is possible. From figure (c), which correspond to CDDW state nodal point excitations with or without tunneling, it may be seen that the some points in the momentum space are ungapped and therefore Fermi arcs are possible. However, in (d) there is a wide gap at the anti-node. The holes are conductive for (c) and (d). The parameter values are, µ = -0.035, $Q_1$ = 0.7742.pi, $Q_2$ = 0.2258.pi, $t = 1$, $t' = -0.12$, $t'' = 0.01$, $t_0 = 0.005$, and $\Delta_0^{PG} = 0.01$.

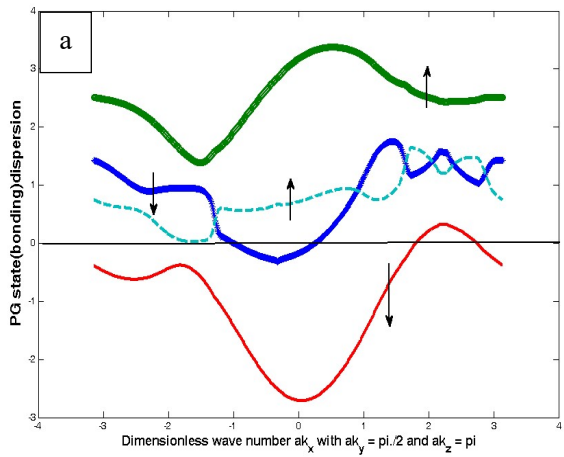
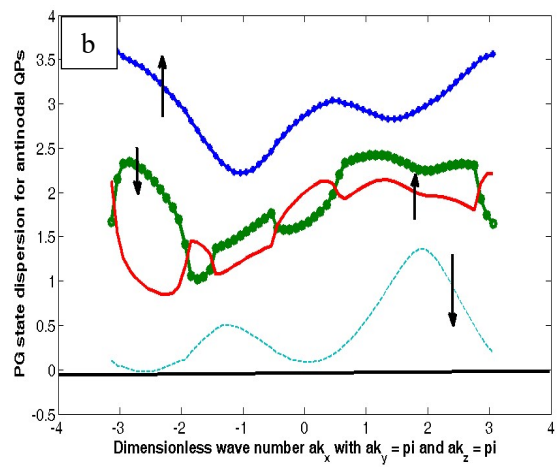

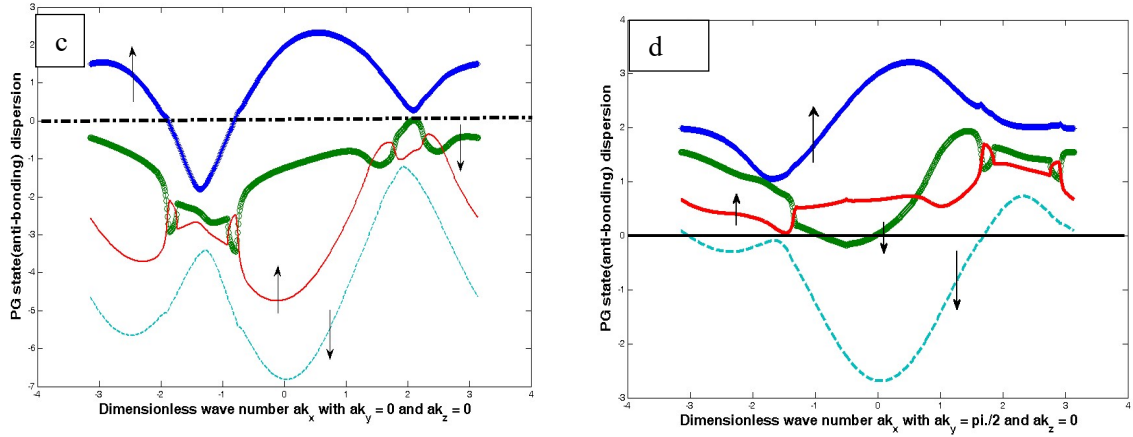

**Figure 2. (a)** A plot of quasi-particle excitation (QP)spectrum given by Eq.(8) in the bonding case as function of dimensionless momentum $k_x a$ for $k_y a = \pi/2$ and $k_z a = \pi$. The horizontal line represents the Fermi energy. Since spin-down valence and conduction bandsare partially empty, the spin-down QP conduction is possible.**(b)** A plot of QPsin the bonding case, as a function of dimensionless momentum $k_x a$ for $k_y a = \pi$ and $k_z a = \pi$ .**(c)** A plot of quasi-particle excitations in the anti-bonding case as function of $k_x a$ for $k_y a = 0$ and $k_z a = 0$. Since the spin-upconduction band is partially empty, the spin-up electron conduction is possible.**(d)** Aplots of QPs in the anti-bonding case as function of dimensionless momentum $k_x a$ for $k_y a = \pi/2$ and $k_z a = 0$. Since the spin-down valence band is partially empty, the spin-down hole conduction is possible.The numerical values of the parameters to be used in the calculation are $t = 1$ , $t'/t = -0.28$, $t''/t = 0.1$, $t'''/t = 0.06$, $t_b/t = 0.3$, $t_z/t = 0.1$, $\frac{\alpha_0}{t} = 0.53$, $\frac{\Delta_0^{PG}}{t} = 0.01$, and $a_0 = 0.4$.

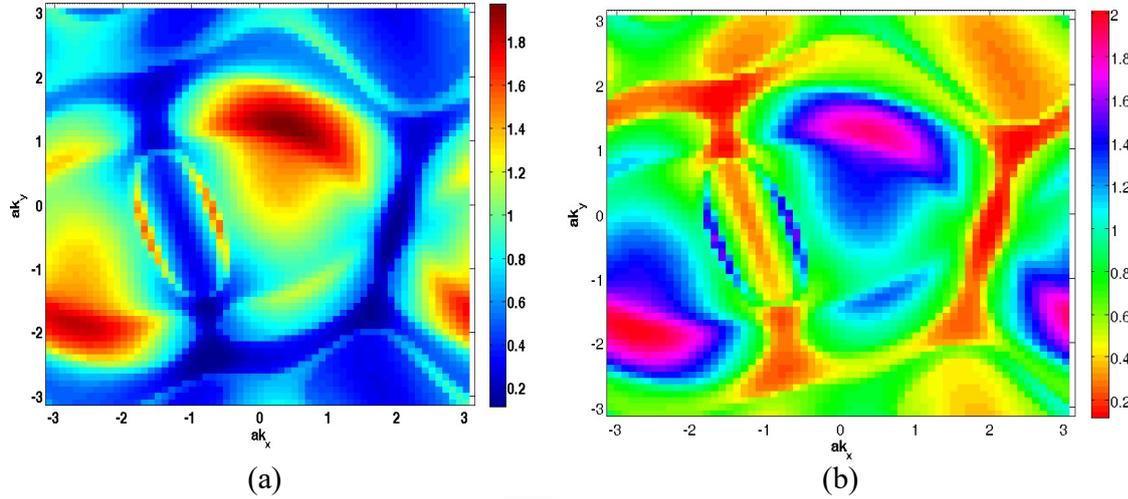

**Figure 3.** The contour plots of the term $\sqrt{\frac{z_0(k)}{2}}$ in the bonding (a) and the anti-bonding (b) cases. The numerical values of the parameters to be used in the calculation are $t = 1$ , $t'/t = -0.28$, $t''/t = 0.1$, $t'''/t = 0.06$, $t_b/t = 0.3$, $t_z/t = 0.1$, $\frac{\alpha_0}{t} = 0.53$, $\frac{\Delta_0^{PG}}{t} = 0.01$, and $a_0 = 0.4$.

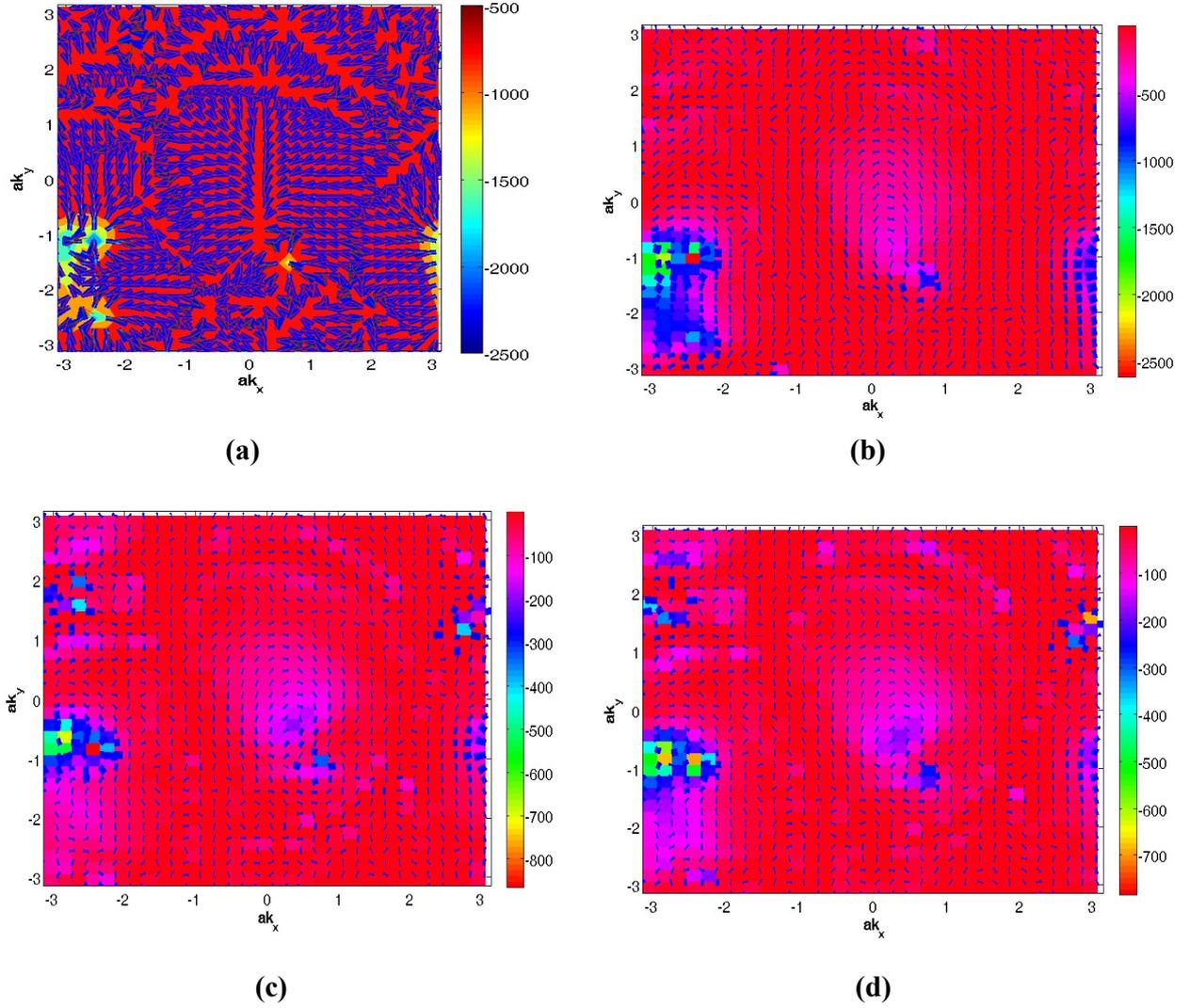

**Figure 4.** Contour plots of spin textures $s_z(n,\mathbf{k})$. (a) $a_0 = 0.40$ and $\alpha_0 = 0.35$. (b) $a_0 = 0.40$ and $\alpha_0 = 0.35$ with a different choice of color and length of arrows. (c) $a_0 = 0.40$ and $\alpha_0 = 0.65$. (d) $a_0 = 0.60$ and $\alpha_0 = 0.65$. The numerical values of the other parameters used in the calculation are t = 1, t′/t = −0.28, t ″/t = 0.1, t ‴/t = 0.06, $t_b/t$ = 0.3, $t_z/t$ = 0.1, $\frac{\Delta_0^{PG}}{t} = 0.01$, μ = 0.00, $Q_1$ = 0.7742.*pi, and $Q_2$ = 0.2258.*pi. The band index 'n' stands for the spin down bands

$$\in (\downarrow, \sigma = \pm 1, k)$$

$$= -\sqrt{\frac{z_0(k)}{2}} + \varepsilon_k^U$$

$$\pm \left( b_0(k) - \left(\frac{z_0(k)}{2}\right) - c_0(k)\sqrt{\frac{2}{z_0(k)}} \right)^{\frac{1}{2}} \text{ intersecting with Fermi energy.}$$

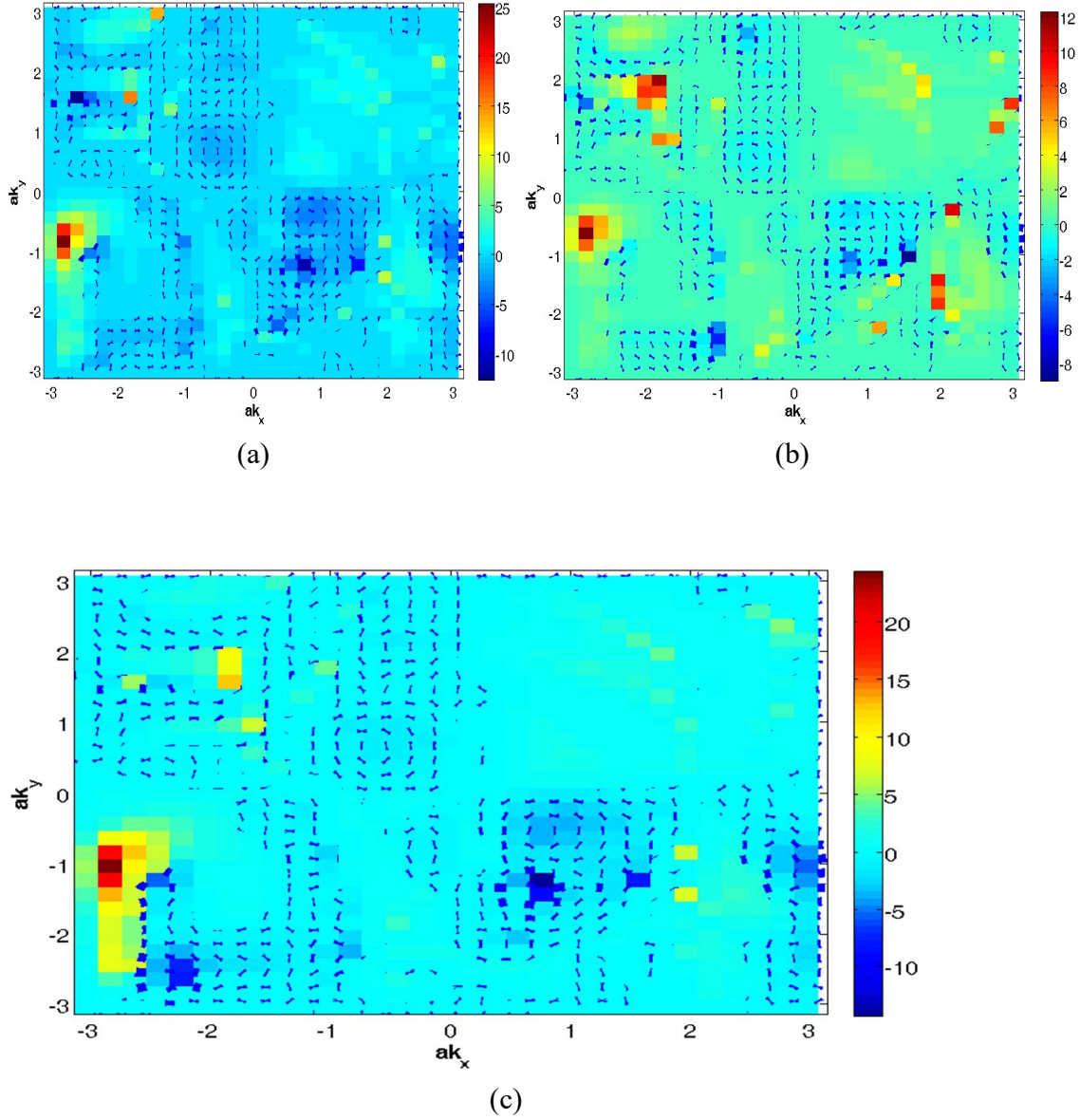

**Figure 5.** Contour plots of the expectation value ($\langle n|S_x n\rangle - \langle n|S_y|n\rangle$). (a) $a_0 = 0.40$ and $\alpha_0 = 0.50$. (b) $a_0 = 0.40$ and $\alpha_0 = 0.70$. (c) $a_0 = 0.70$ and $\alpha_0 = 0.50$. The numerical values of the other parameters used in the calculation are t = 1, t'/t = −0.28, t''/t = 0.1, t'''/t = 0.06, $t_b/t$ = 0.3, $t_z/t$ = 0.1, $\frac{\Delta_0^{PG}}{t} = 0.01$, μ = 0.00, $Q_1$ = 0.7742.*pi, and $Q_2$ = 0.2258.*pi. The bands $\in (\downarrow, \sigma = \pm 1, k) = -\sqrt{\frac{z_0(k)}{2}} + \varepsilon_k^U \pm \left(b_0(k) - \left(\frac{z_0(k)}{2}\right) - c_0(k)\sqrt{\frac{2}{z_0(k)}}\right)^{\frac{1}{2}}$, intersecting with Fermi energy, are chosen for the purpose of the plot.